\documentclass[aps,prl,twocolumn,superscriptaddress]{revtex4}
\usepackage{latexsym}
\usepackage{graphicx}
\usepackage{epsfig}
\newcommand \be {\begin{equation}}
\newcommand \bea {\begin{eqnarray}}
\newcommand \ee {\end{equation}}
\newcommand \eea {\end{eqnarray}}
\newcommand \bed {\begin{displaymath}}
\newcommand \eed {\end{displaymath}}

\newcommand{\bit}{\begin{itemize}}

\begin{document}

\title{Nucleon correlations  and Higher Twist effects in  nuclear structure functions}
\author{Paolo Castorina}
\affiliation{Dept. of Physics, University of Catania and INFN, Sezione di Catania
\quad \\
Via Santa  Sofia 64, I-95123, Catania, Italy
}
\date{\today}
\begin{abstract}

The overlap of the nucleons in nuclei plays an important role in understanding the nuclear dependence of deep inelastic scattering data.
It is shown that the nuclear modification of the higher twist  scale can be essentially determined by the overlapping volume
 per nucleon and this effect gives a large contribution to nuclear shadowing for small $x$ and low $Q^2$.
In this kinematical region there is also a moderate enhancement of the longitudinal
structure function in nuclei.

\end{abstract}

\pacs{............................}

\maketitle

The interpretation of the HERA and RHIC data and the future experiments at LHC require a deep understanding of the behaviour of the parton distributions in the 
small $x$ region both for the free nucleon and for nuclei. 

The main ideas of the  dynamical description of the small $x$ physics are the saturation of the parton densities  and the shadowing effects.
Both these problems can be studied from the fundamental point of view by the non linear QCD evolution equations \cite{nlee}. However, since they are related with 
confinement, some complementary approaches,as classical gluon field theory \cite{mc} or the pomerons exchange \cite{sandy}
,  take into account the non perturbative dynamics.

Nuclear shadowing is a crucial ingredient of the initial state effects in hard phenomena in proton-nucleus and nucleus-nucleus collisions and then  some useful
, model independent, phenomenological analyses of the nuclear modifications of the parton distributions have
been proposed \cite{eks}.   

The major difficulty for a clear theoretical description of the nuclear shadowing is that the available data on deep inelastic scattering (DIS) 
on nuclear target in the small $x$ region are also 
for small $Q^2$ and then the standard QCD  evolution becomes unreliable.

Recently the calculation of the nuclear parton distribution functions  in the Gribov approach \cite{frank1} has been updated \cite{frank2}
 and the failure of the leading twist model to describe the DIS data  indicates significant higher twist (HT) contributions in the small
$x$  region. In particular, the HT contribution to nuclear shadowing turns out about $ 50 \%$ of the whole suppression and the authors
 argue that in the kinematical range covered by data it is possible to have this large corrections due to diffractive effects. On the other hand, in ref. 
\cite{nestor} the HT contribution turns out small ( see later).

In this letter I suggest  that in the small $x$ and low $Q^2$ region the HT contributions are relevant for the agreement 
with the present day data on nuclear shadowing.
This conclusion, that can be also obtained by resumming QCD power corrections \cite{qiuvitev},
is based on  an intuitive, coarse-grain analysis of 
the HT effects  by  the overlap of nucleons in the nucleus.

When nucleons overlap, the effective average volume per nucleon decreases due to nuclear density and correlations.
 This reduction of the average interquark distances
, due to overlapping volume per nucleon $V^{op}_A$ in the nucleus $A$, is an effective way to describe a greater number of quarks in a larger bag with
, however, a  reduced effective volume per nucleon, $V_A$, with respect to the free nucleon volume $V_N$.

This idea has been applied to  the EMC effect
 ,by the  $Q^2$-rescaling \cite{bob} model, and also to the nuclear shadowing  \cite{noi}.

 I shall show that by a simple analysis, based on the overlapping volume per nucleon in the nucleus, one can suggest that:

i) there is a nuclear modification of the HT scale which gives a relevant contribution to nuclear shadowing in the present day data;

ii) the nuclear correction to the longitudinal structure function, $F_{2A}^L$, are not so large as preliminary reported in ref. \cite{ack}.

Let us first discuss the higher twist effects  which  are subasymptotic scaling violations, $O(\mu^2/Q^2)$,
 whose typical scale $\mu^2$ is fixed by the matrix elements 
of the relevant operators in the light-cone expansion of the product of the currents operators.In particular, Jaffe and Soldate \cite{jaffe} 
found a complete set of nine operators, totally symmetric, traceless in Lorentz indices and depending on color, spin and flavor indices, which permits us to evaluate 
the twist four, spin two, contribution to electroproduction in QCD.

The HT formalism has been largely applied \cite{mulders} and it  requires a specific model to evaluate the non perturbative, target dependent, matrix elements.
For example, in the MIT bag model the twist four correction to the nucleon structure function $F_{2N}$ turns out
\be
\int dx F_{2N}^{HT} = \frac{1}{2} \frac{\pi \alpha_c}{Q^2 M_N V_N} ( -\frac{131}{27} K_1 + 8 K_2)
\ee  

where $M_N$ and $V_N$ are respectively the nucleon mass and volume, $\alpha_C$ is an effective coupling constant in the bag ( usually $\alpha_c=0.5$ is assumed) 
and $K_1=1.08$ and $K_2=0.17$ are integrals over $V_N$ of the bag model spinors for massless 
quarks in the lowest modes \cite{jaffe}.

This result, by itself, is  not useful for the comparison with DIS data because one needs to analyze the $x$ dependence and, moreover,in the small $x$, low $Q^2$
region the valence approximation of the bag model is not reliable.A complete calculations of the matrix elements is still missing and, on the other hand,
 it is quite useful to have, at least, a phenomenological understanding of the HT contributions, crucial for the agreement with data at small $Q^2$.

Then, following the structure of eq.(1), let us write 
\be
 F_{2N}^{HT} = \frac{1}{2} \frac{\pi \alpha_c}{Q^2 M_N V_N} C_N(x)
\ee  
 and try to have some  indication of the effective weight of the matrix elements  for determining the HT scale of the nucleon in the small $x$ region.
The next step will be to generalize the analysis for  a nuclear target.

However, before entering in the phenomenological details, let us consider some results which follow from  general arguments. 

Let us write the $F_2$ structure functions as a leading twist term (LT) plus a  $O(1/Q^2)$ HT contribution. For the nucleon case we shall approximate  $F_{2N}= F_{2D}/2$
since at high energy the nuclear effects in deuterium are negligible and from now  on the formulas refere to structure functions per nucleon.For deuterium one has 
$ F_{2D} = F_{2D}^{LT}+ F_{2D}^{HT}$
   and for a larger nucleus $A$,
 $F_{2A} = F_{2A}^{LT}+ F_{2A}^{HT}$.
By defining 
 $R_A^{LT}= {F_{2A}^{LT}}/{ F_{2D}^{LT}}$
and $ R_A= {F_{2A}}/{ F_{2D}}$
  ,to order  $O(1/Q^2)$, one obtains 
\be
 R_A [1- \frac{F_{2A}^{HT}}{ F_{2A}}+ \frac{F_{2D}^{HT}}{ F_{2D}}] = R_A^{LT}
\ee
Then the HT corrections can suppress the (LT) term only if  
 $ R_A F_{2D}^{HT} > F_{2A}^{HT}$ .Since in the small $x$ region $R_A<1$, if the HT correction increases the (LT) 
term (i.e. it is  positive) then the effective HT scale must be larger in deuterium with respect to 
the larger nucleus $A$ . Viceversa, if the HT term are negative the larger nuclei have a larger effective scale at small $x$.

Concerning the sign of the  HT correction to $F_{2D}$ let us notice that:
1) In the MIT bag model this overall correction is negative ( see eq. (1));
2) in the parametrization of the deuteron structure function data in ref. \cite{nm} the  $O(1/Q^2)$ leading correction in the small $x$ region is negative.

By assuming this  indication, one concludes that the nuclear HT corrections are negative and there is a larger HT effective scale at small $x$ with respect to 
the nucleon case.Moreover one can use the phenomelogical input given in ref. \cite{nm} and write  that
\be
 F_{2D}^{HT} = \frac{1}{2} \frac{\pi \alpha_c}{Q^2 M_N V_N} C_D(x) \simeq - \frac{c_1 x+ c_2 x^2}{Q^2}  
\ee  
where $c_1 = 1.509 Gev^2$ and  $c_2 = -8.553 Gev^2$ \cite{nm}.Of course  this  approximation 
is valid only in the small $x$  region considered in the data of ref.  \cite{nm}.

Let us now consider the nuclear effects.
The HT corrections come essentially from the matrix elements of quark-quark-gluon and/or four quark operators \cite{jaffe} in the target and then 
in a nucleus there are peculiar contributions, with respect to the free nucleon, due to the nuclear binding and to partons coming from different
 nucleons in the nucleus. A rigorous calculation is quite hard, but, according to  the parametrization in eq.(2), it is natural a description
 of the nuclear effects which takes into account the nuclear binding energy by an effective mass of the nucleon, $M^*=M_N+\epsilon$ 
(where $\epsilon$ is the removal energy \cite{aku},and the correlation
among nucleons by their overlap in the nucleus as previously  discussed. 
  Within this extremely simplified approach of the nuclear effects, the HT contribution to $F_{2A}$ per nucleon can be parametrized as ( see eq.(2))
 \be
 F_{2A}^{HT} = \frac{1}{2} \frac{\pi \alpha_c}{Q^2 M^* V_A} C_A(x)
\ee  
where  $V_A/V_N= 1-V^{op}_A /V_N$ and $C_A(x)$, related to the matrix element of the relevant operator in the nucleus, takes also into account the
 HT contribution due to partons from different nucleons.It is reasonable to assume that this dynamical effect is proportional 
to the overlapping volume per nucleon, $V^{op}_A$, and then to write
\be
 C_A(x)= C_D(x)+C(x) \frac{V^{op}_A}{V_N}
\ee  
where $C(x)$ is independent on $A$.

Then the HT nuclear correction  turns out that
\be
 F_{2A}^{HT} =  F_{2D}^{HT} \frac{M_N}{M^*} \frac{V_N}{V_A} [ 1+ \gamma(x)\frac{V^{op}}{V_N}] \equiv F_{2D}^{HT} \Sigma_A
\ee  
where $\gamma(x)=C_A(x)/C_N(x)$  will be approximated with a constant parameter, $\gamma$, in such a way that the HT nuclear scale 
 is completely determined by the overlapping volume per nucleon. 
By combining the previous eqs.(3,7), the relation between $R_A^{LT}$ and $R_A$ in terms of the HT corrections is given by
\be
 R_A ={{R_A^{LT} +\frac{F_{2D}^{HT}}{F_{2D}} \Sigma_A} \over {1 + \frac{F_{2D}^{HT}}{F_{2D}}}}
\ee
which is the starting point for the comparison with the experimental data.

In the previous formula the only free parameter is $\gamma$.
Indeed the overlapping volume per nucleon  has been evaluated in ref. \cite{bob} by including a Reid-soft core potential  between nucleons 
and considering only the two nucleon overlap. In ref. \cite{vedi} the previous evaluation has been reanalized with a reduction of a factor 
$\simeq 0.65$ with respect to the initial values. Then the overlapping volume per nucleon for different nuclei is fixed as the $65 \%$
  \cite{acc} 
of the values obtained in ref. \cite{bob}.
The removal energy $\epsilon$ is small ( few Mev) and negative and  it has a negligible quantitative effect.
 $F_{2D}^{HT}$ is given in eq.(4) and  $F_{2D}$ is experimentally  known for small value of $x$ and accurately parametrized 
\cite{nm}.
Finally, one has to know $R_A^{LT}$ and different LT inputs change
the fitted value of the parameter $\gamma$.

From this point of view, it should be clear that the agreement with data depends on the relative weight between the LT and the HT contributions.
The power corrections are an important element of the whole analysis but their magnitude is correlated with the  
the initial parametrization of the structure functions for the $Q^2$ evolution. In ref. \cite{frank2} a model of diffraction 
is used in order to obtain an initial condition for DGLAP evolution at $Q_O^2 = 4$ $Gev^2$. On the contrary, in ref.\cite{nestor} 
a model, to all twist orders, is proposed for the full low $Q^2$ region
and there is no QCD evolution.
In this latter framework, where a small HT correction is reported, is not simple to handle the $O(1/Q^2)$ and 
clearly separate a ``LT'' term. Then, in the present analysis,
one considers  the leading twist results in the Gribov approach obtained in ref. \cite{frank2}. 
In Fig.1 is reported the comparison of the leading twist theory plus higher twist contributions , according to eq.(8)
to the NMC data \cite{nmc1}on $F_{2C}/F_{2D}$.The dashed and dotted-dashed lines are respectively for  $\gamma=3$ and $\gamma=4$ and
the dotted line is the leading twist result in ref.\cite{frank2}.
 Fig.2 is the same comparison with the Ca-40 data and  in Fig.3 the comparison is with the ratio $F_{2PB}/F_{2C}$.
In fig.4 has been considered the $Q^2$ dependence at fixed $x=0.0125$ of the ratio $F_{2Sn}/F_{2C}$ \cite{nmc11}.

\begin{figure}
\epsfig{file=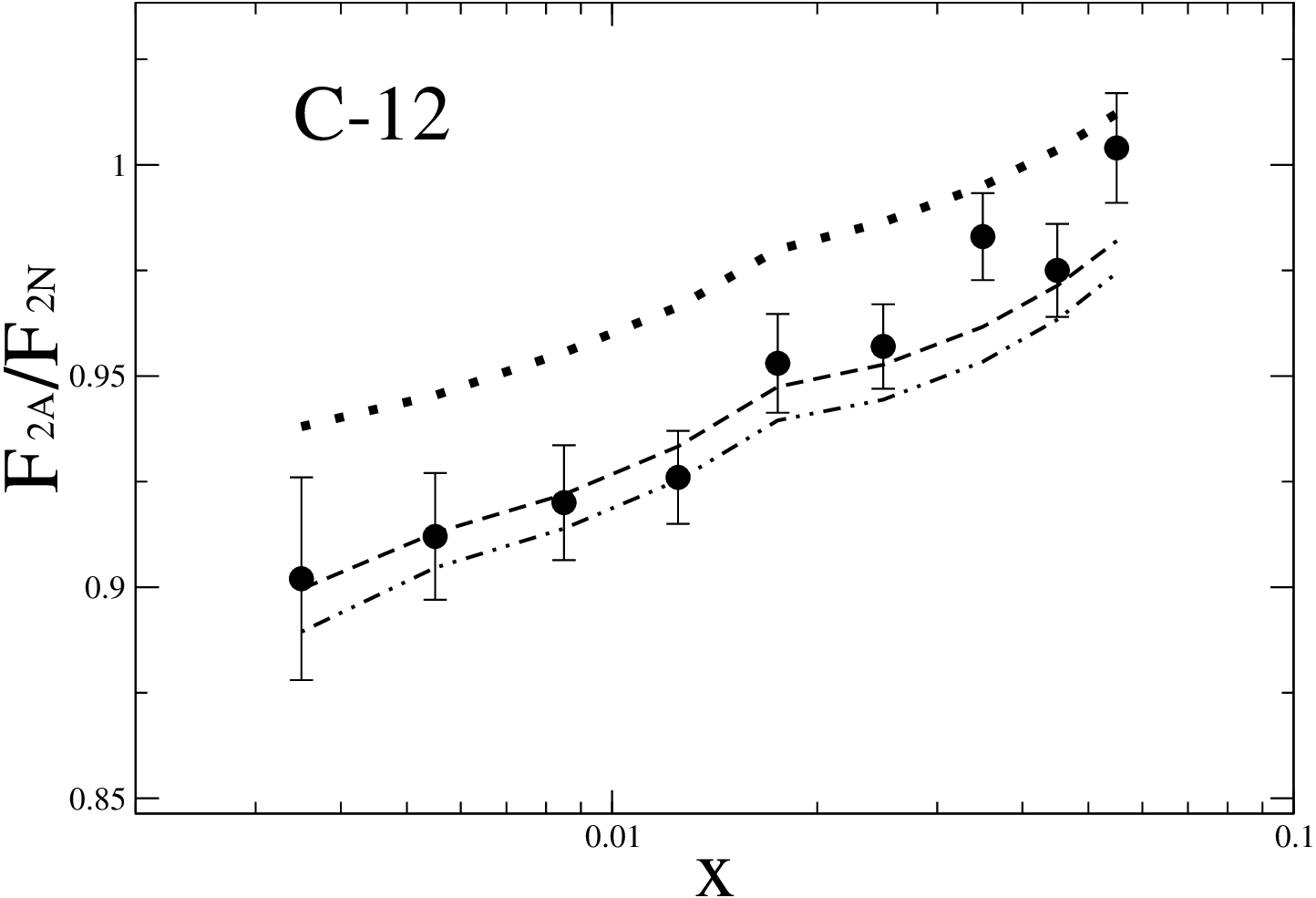,height= 6.0 true cm,width=6.0 true cm, angle=0}
\caption{Fig.1  Comparison of the leading twist theory plus higher twist contributions  according to eq.(8)
to the NMC data \cite{nmc1}on $F_{2C}/F_{2D}$, sees text.The statistical and systematic  errors are quadratically added.}
\end{figure}

\begin{figure}
\epsfig{file=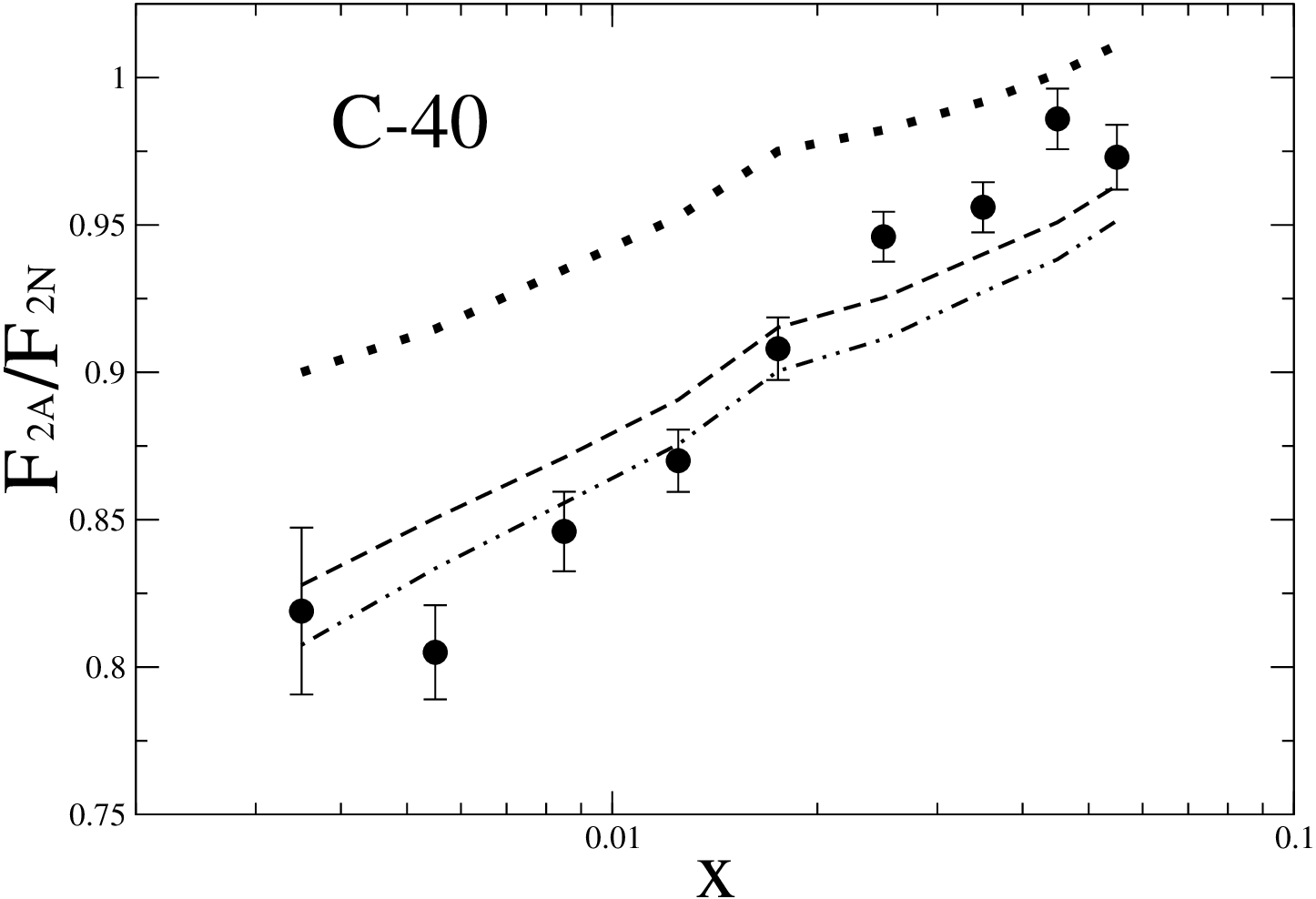,height= 6.0 true cm,width=6.0 true cm, angle=0}
\caption{Fig.2  Comparison of the leading twist theory plus higher twist contributions , according to eq.(8)
to the NMC data \cite{nmc1}on $F_{2Ca}/F_{2D}$, see text.The statistical and systematic errors are quadratically added.}
\end{figure}

\begin{figure}
\epsfig{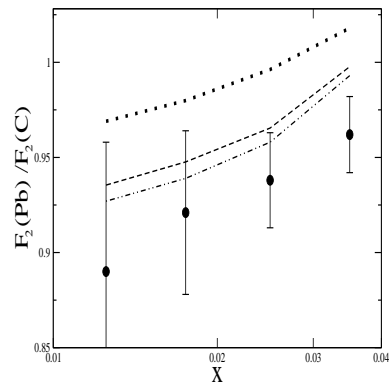}
\caption{Fig.2  Comparison of the leading twist theory plus higher twist contributions 
to the NMC data on  $F_{2Pb}/F_{2C}$,see text.The statistical and systematic errors are quadratically added.}
\end{figure}

\begin{figure}
\epsfig{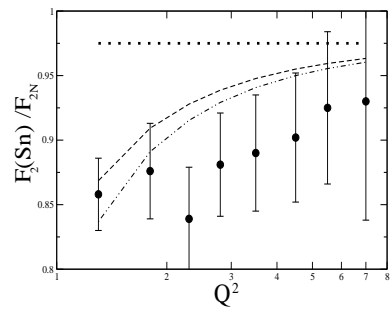}
\caption{Fig.2  Comparison of the leading twist theory plus higher twist contributions 
to the NMC data on  $F_{2Sn}/F_{2C}$ at $x=0.0125$ and different $Q^2$,see text.
The statistical and systematic errors are quadratically added.
 The dotted line is the leading twist 
result $0.975$ \cite{frank2}.}
\end{figure}

Let us now comment on the longitudinal structure function $F_{2A}^L$.
In ref. \cite{ack} has been reported a quite large value ( about a factor 4) of the ratio between the longitudinal and transverse structure 
functions in N-14 for  $x \simeq 0.0045$ and  low $Q^2$ with respect to the deuterium case.

This ehancement has been investigated by many authors \cite{molti,qiuvitev,mio} with the conclusion that a large HT correction 
for nytrogen 
is required.
Indeed, also in this case, the effect is due to a balance between LT term and HT correction 
which  depends on the absolute value of the HT scale
rather than on the  ratio between the nitrogen and deuterium HT scale. 
By the present analysis,  an enhancement should be found because the  HT nuclear scale increases. However 
 it turns out that   
the effective HT nytrogen scale at $x \simeq 0.0045$ is much smaller than the the value of $0.056 Gev^2$ required \cite{mio} to obtain the reported large correction. 

Moreover  one has to remember that:1) the matrix elements which produce the HT correction for the $F_L$ and $F_2$ structure functions 
are correlated but  are not  the same 
and then the previous results on the HT scale in nuclei cannot be directly applied to the longitudinal structure function;
2) the data in ref. \cite{ack} are under reanalysis  \cite{veditu}.

Let us conclude with  few comments.

The question if the observed nuclear shadowing is a leading twist or a higher twist effect is still open.

In the present analysis, the HT contribution to nuclear shadowing is relevant, according to  ref. \cite{frank2},
but this could depend on the initial parametrization.
 A conclusive answer requires an updating of the phenomenological fits
which includes the $O(1/Q^2)$ corrections, the QCD evolution and  also the experimental data on the longitudinal structure
 function.

The overlap of the nucleons is an important element in understanding the nuclear shadowing ( also at  leading twist \cite{noi})  
and the EMC effect \cite{bob}.
This can be probably considered as a first  signal towards a geometrical critical transition in heavy ions collision at high energy and density,
associated with percolation \cite{helmut} 
rather than  with the temperature.

The author thanks Nestor Armesto, Carlos Salgado and U.Wiedemann for very useful suggestions
 and the CERN Theory Division for hospitality.

\end{document}